\documentclass[journal]{IEEEtran}

\IEEEoverridecommandlockouts
\usepackage[T1]{fontenc}
\usepackage{algorithmic}
\usepackage[ruled,vlined]{algorithm2e}
\usepackage{enumitem}
\usepackage{graphicx}
\usepackage{amsmath,amssymb,amsfonts}
\usepackage{array}
\usepackage[font=small]{caption}
\usepackage{paralist}
\usepackage{mathtools}
\usepackage{lettrine}
\usepackage{cite}
\usepackage{xcolor}
\usepackage{cancel}
\usepackage[caption=false,font=footnotesize]{subfig}

\def\BibTeX{{\rm B\kern-.05em{\sc i\kern-.025em b}\kern-.08em  T\kern-.1667em\lower.7ex\hbox{E}\kern-.125emX}}
\usepackage{tikz,pgf} \usetikzlibrary{intersections,calc} \usetikzlibrary{arrows.meta,patterns}

\begin{document}

\title{Wideband Effects on Near-Field Pose Estimation of Target-Lodged RIS
\thanks{This work was partially supported by the European Union under the Italian National Recovery and Resilience Plan (NRRP) of NextGenerationEU, partnership on “Telecommunications of the Future” (PE00000001 - program “RESTART”).} 
}

\author{Dario~Tagliaferri,~\IEEEmembership{Member,~IEEE}}

\maketitle

\begin{abstract}
Reconfigurable intelligent surfaces (RISs) have been recently considered in sensing context to either improve localization performance or to extend the coverage in non-line-of-sight scenarios. While most of the literature considers RISs as statically placed in the environment, the usage of target-lodged RISs is relatively new and could be of interest where the target's reflectivity can/must be increased to improve its detection or parameters' estimation. This letter derives the Cramér-Rao bound (CRB) on the estimation of position and orientation (\textit{pose}) of a target-mounted RIS, in generic conditions: near-field, bistatic and wideband operation (i.e., when the wavefront across the RIS is curved and the employed sensing bandwidth is large enough to obtain a frequency-dependent RIS behavior). In particular, we focus on the wideband effect, that implies a \textit{pose-dependent filtering} on the impinging signal decreasing or increasing the CRB depending on the RIS size and the employed signal bandwidth. 
\end{abstract}

\begin{IEEEkeywords}
Cramér-Rao bound, near-field, wideband, position, orientation, RIS  
\end{IEEEkeywords}

\section{Introduction}\label{sect:intro}
Electromagnetic (EM) metasurfaces are rapidly surging in both academia and industry as one of the promising technologies for the smart propagation in the next generation of communication systems (6G) \cite{di2020smart}. In particular, reconfigurable intelligent surfaces (RISs) are discrete metasurfaces made of passive regularly spaced, sub-wavelength-sized elements, whose reflection coefficient can be dynamically tuned to manipulate the direction of reflection and refraction of an impinging wave \cite{zhang2021performance}. 

The usage of RIS for sensing is relatively recent  \cite{9775078}. In~\cite{Buzzi_RISforradar_journal}, the authors proposed to use a RIS to assist radar in non-line-of-sight conditions, focusing on the maximization of the probability of correct detection of a given target, via signal-to-noise (SNR) maximization with constrained phase design. The results analyze the Cramér-Rao bound (CRB) on position estimation and suggest the placement of the RIS nearby either the radar or the target. The work \cite{Zhang2022metalocalization} considers indoor localization by means of wall-placed RIS. The paper \cite{Wang2022_location_awareness} compares discrete and continuous metasurfaces in terms of CRB for position estimation in the integrated sensing and communication context. With the advent of large RISs, that enable ultra-high precision localization, the near-field (NF) propagation condition applies, i.e., the wavefront across the RIS is no more flat and allows proper energy focusing through reflection. NF operation opened for further research on fundamental localization bounds and practical RIS phase configuration. The work \cite{9625826} derives the CRB for position and orientation estimation in some selected practical cases, assuming NF operating conditions and discussing the phase configuration. The authors of \cite{9650561} propose a practical NF RIS phase configuration and analyze the CRB on position estimation. 

In all the the aforementioned literature, RISs are statically placed in the environment and used to aid localization, by magnifying the Tx/Rx aperture to decrease the CRB. In some scenarios, however, the RIS can be directly lodged on targets, to enhance their "visibility" in the sensing data and improve their localization performance. This setup is only considered in \cite{Zhang2023_targetRIS} and \cite{10001209}. In the former, the authors investigate the CRB on position and orientation estimation for a target-mounted RIS under far-field (FF) assumption, while the latter addresses the CRB evaluation on position estimation with a perfect RIS phase configuration and bistatic settings. However, none of the previous works evaluate the \textit{wideband} effects, namely the change of the RIS response due to a change in the frequency within the employed signal bandwidth. The only work addressing this issue is \cite{Wymeersch2022_wideband_RIS}, for a single-input-single output system and RIS again placed in the environment. In FF conditions, the wideband effect gives rise to \textit{reflection beam squinting}, i.e, a frequency-dependent spatial spreading of the reflected signal around the desired direction diminishing the amount of received energy. Similarly, a RIS under NF and wideband operation is subject to \textit{defocusing}.  

This letter tackles the problem of pose (3D position plus 3D orientation) estimation of a target-lodged RIS, in a generic  bistatic, NF and wideband (frequency-selective) setting. We derive the CRB for perfect RIS phase configuration (although imperfect phase configuration can be considered as well). Further, we outline the importance of considering the wideband RIS behavior in practical cases, where the RIS acts as a \textit{pose-dependent filter} for the impinging sensing signal. Remarkably, the latter effect allows increasing or reducing the available information w.r.t. the narrowband case (where the RIS is assumed to behave like a frequency-independent target) depending on the RIS size and the specific employed bandwidth value.  

The letter is organized as follows: Sect. \ref{sect:system_model} presents the system model, in Sect. \ref{sect:CRB} we evaluate the CRB, Sect. \ref{sect:results} presents the numerical results while Sect. \ref{sect:conclusion} concludes the paper. We adopt the following notation: bold upper- and lower-case denote matrices and column vectors. The L2-norm of a vector is denoted with $\|\cdot\|$. Matrix transposition and conjugate transposition are indicated respectively as $\mathbf{A}^T$ and $\mathbf{A}^H$. $\mathbf{I}_n$ is the identity matrix of size $n$.  $\mathbf{a}\sim\mathcal{CN}(\boldsymbol{\mu},\mathbf{C})$ denotes a multi-variate circularly complex Gaussian random variable with mean $\boldsymbol{\mu}$ and covariance $\mathbf{C}$. $\mathrm{trace}(\mathbf{A})$ extracts the trace of $\mathbf{A}$. $\mathbb{R}$ and $\mathbb{C}$ stand for the set of real and complex numbers, respectively. $\delta_{n}$ is the Kronecker delta.

\section{System Model}\label{sect:system_model}

\begin{figure}
    \centering
    \includegraphics[width=\columnwidth]{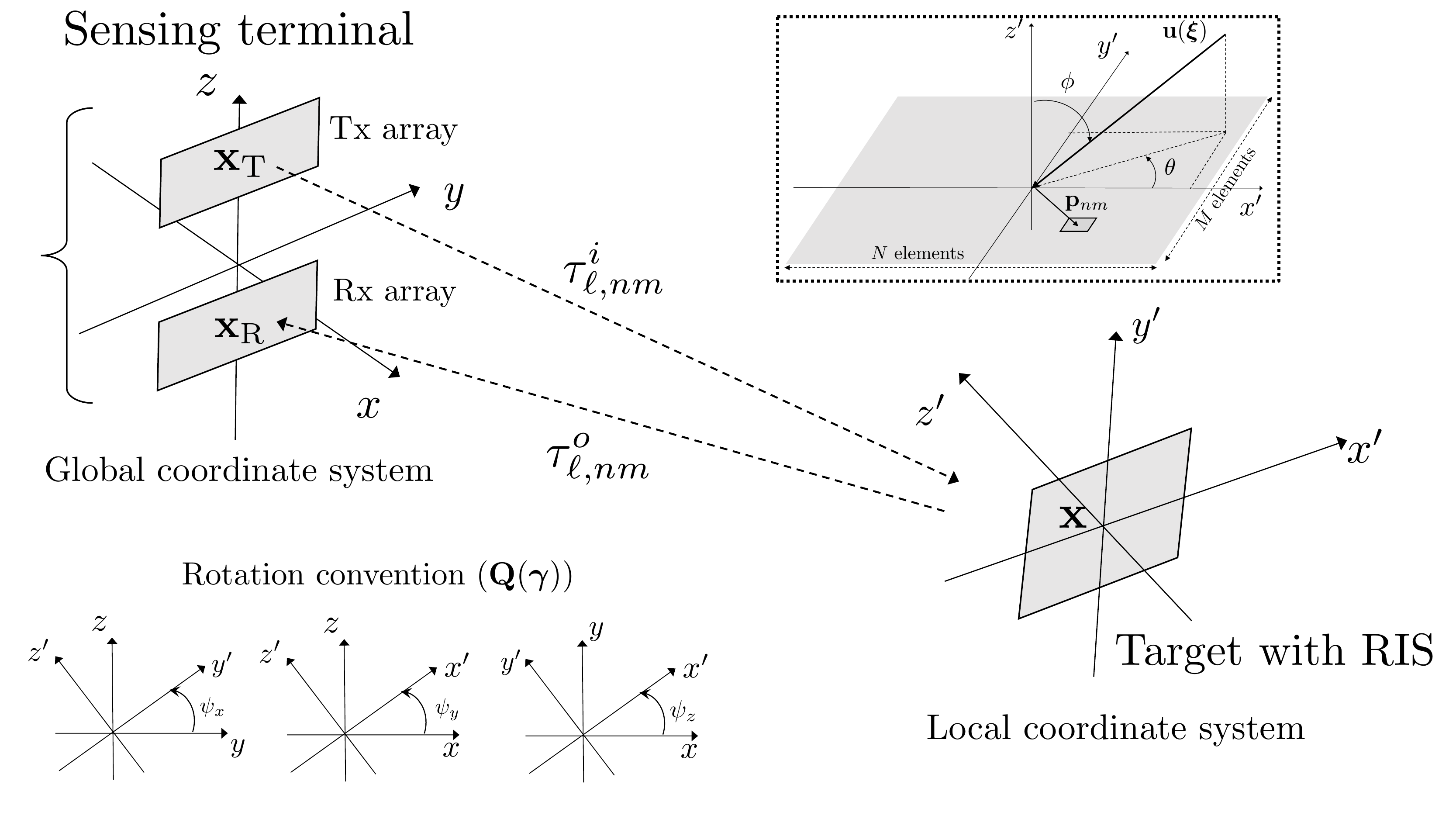}
    \caption{System model.}
    \label{fig:system_model}
\end{figure}
Let us consider a generic sensing terminal in a bistatic setting (e.g., a multiple-input multiple-output (MIMO) radar) that aims at estimating the position and orientation of a target equipped with a RIS (Fig. \ref{fig:system_model}). In a \textit{global} reference system, the phase centers of transmitting (Tx) and receiving (Rx) antenna arrays are located in
$\mathbf{x}_\mathrm{T}\in\mathbb{R}^{3\times 1}$ and $\mathbf{x}_\mathrm{R}\in\mathbb{R}^{3\times 1}$, respectively. The sensing terminal implements $L$ measurement channels, for all Tx-Rx antenna pairs (say, each single antennas is located in $\mathbf{x}_{\mathrm{T},\ell}$ and $\mathbf{x}_{\mathrm{R},\ell}$, respectively). At each Tx antenna, the sensing terminal emits the signal
\begin{equation}\label{eq:TX_signal}
    s(t) = g(t) e^{j2 \pi f_0 t}
\end{equation}
where the base-band waveform $g(t)$ has bandwidth $B$ and the carrier frequency is $f_0$. For simplicity, we assume $g(t)$ to be the same on each Tx antenna, as for a MIMO radar operating in time-division multiplexing. 
The RIS is in the generic position $\mathbf{x}\in\mathbb{R}^{3\times 1}$ and is made by $N \times M$ elements, displaced along the $x$ and $y$ axes of the of the \textit{local} reference system. The local axes are rotated w.r.t. the global ones by the Euler angles $\boldsymbol{\gamma}=[\psi_x, \psi_y, \psi_z]^T\in\mathbb{R}^{3\times 1}$ (roll $\psi_x$, pitch $\psi_y$ and yaw $\psi_z$). These latter represent the RIS orientation, that is the objective of the estimation together with position $\mathbf{x}$. The position of the $(n,m)$-th element of the RIS in global coordinates is  
\begin{align}
    \mathbf{x}_{nm} & = \mathbf{x} +  \mathbf{Q}(\boldsymbol{\gamma})\mathbf{p}_{nm} 
\end{align}
for $n=-N/2,...,N/2-1$, $m=-M/2,...,M/2-1$, where $\mathbf{p}_{nm} = \left[nd, \,md,\,0\right]^\mathrm{T}$ is the local position of the $(n,m)$-th element in local coordinates (for inter-element spacing $d$), matrix $\mathbf{Q}(\boldsymbol{\gamma})\in \mathrm{SO}(3) \overset{\triangle}{=}\left\{\mathbf{Q}| \mathrm{det}(\mathbf{Q})=1, \mathbf{Q}\mathbf{Q}^T = \mathbf{I}_3\right\}$ defines the relative counterclockwise rotation of the local reference system around $z$, $y$ and $x$ axis by angles $\psi_z$, $\psi_y$, $\psi_x$, respectively. The expression of $\mathbf{Q}(\boldsymbol{\gamma})$ is:
\begin{equation}
    \mathbf{Q}(\boldsymbol{\gamma}) = \mathbf{Q}_z(\psi_z) \mathbf{Q}_y(\psi_y) \mathbf{Q}_x(\psi_x)
\end{equation}
where single cascaded rotation matrices are defined according to the convention indicated in Fig. \ref{fig:system_model}. 

The Rx signal for the $\ell$-th measurement channel of the sensing terminal (after demodulation) is reported in \eqref{eq:Rx_signal}, 
\begin{figure*}
\begin{equation}\label{eq:Rx_signal}
    \begin{split}
        y_\ell(t) & = \rho \sum_{n,m} e^{j\Phi_{nm}} e^{-j 2 \pi f_0 (\tau^i_{\ell,nm} + \tau^o_{\ell,nm})} g\left(t-\tau^i_{\ell,nm} - \tau^o_{\ell,nm}\right) + z_\ell(t)\\
        & \overset{(a)}{\approx} \rho \, e^{-j 2 \pi f_0 (\tau^i_0+\tau^o_0 + \Delta\tau^i_{\ell} + \Delta\tau^o_{\ell})} \sum_{n,m}e^{j\Phi_{nm}}
        e^{-j 2 \pi f_0 (\Delta\tau^i_{nm}+\Delta\tau^o_{nm})}  g\left(t\hspace{-0.1cm}- \hspace{-0.1cm}\tau^i_0\hspace{-0.1cm}-\hspace{-0.1cm}\tau^o_0\hspace{-0.1cm} - \hspace{-0.1cm}\Delta\tau^i_{\ell}\hspace{-0.1cm} - \hspace{-0.1cm}\Delta\tau^o_{\ell} - \Delta\tau^i_{nm}\hspace{-0.1cm}-\hspace{-0.1cm}\Delta\tau^o_{nm}\right) + z_\ell(t) \\
        & \overset{(b)}{\approx} \rho \, g(t-\tau^i_0-\tau^o_0)\,e^{-j 2 \pi f_0 (\tau^i_0+\tau^o_0 + \Delta\tau^i_{\ell} + \Delta\tau^o_{\ell})} \sum_{n,m}e^{j\Phi_{nm}}
        e^{-j 2 \pi f_0 (\Delta\tau^i_{nm}+\Delta\tau^o_{nm})}+ z_\ell(t)
    \end{split}
\end{equation}
\hrulefill
\begin{equation}\label{eq:Rx_signal_frequency}
    \begin{split}
        Y_\ell(f) & = \rho \,G(f) \sum_{n,m} e^{j\Phi_{nm}} e^{-j 2 \pi (f_0+f) (\tau^i_{\ell,nm} + \tau^o_{\ell,nm})} + Z_\ell(f) \\
        & \overset{(a)}{\approx}  G(f) e^{-j 2 \pi (f_0+f) (\tau^i_0+\tau^o_0 + \Delta\tau^i_{\ell} + \Delta\tau^o_{\ell})}  \beta(f|\mathbf{\Phi}) + Z_\ell(f) \hspace{-0.1cm}\overset{(b)}{\approx} \hspace{-0.1cm} G(f) 
        e^{-j 2 \pi (f_0+f) (\tau^i_0+\tau^o_0)}
         e^{-j 2 \pi f_0 ( \Delta\tau^i_{\ell} + \Delta\tau^o_{\ell})} \beta(0|\mathbf{\Phi}) \hspace{-0.1cm} + \hspace{-0.1cm}Z_\ell(f)
    \end{split}
\end{equation}
\hrulefill
\end{figure*}
where $\rho$ denotes geometrical energy losses, $\Phi_{nm}$ is the phase applied at the $(n,m)$-th element of the RIS, 
\begin{equation}\label{eq:delays}
    \tau^{i}_{\ell,nm} = \frac{ \|\mathbf{x}_{nm}-\mathbf{x}_{\mathrm{T},\ell}\|}{c}, \,\,\,\,\, \tau^o_{\ell,nm} = \frac{\|\mathbf{x}_{\mathrm{R},\ell} - \mathbf{x}_{nm}\|}{c},
\end{equation}
are the absolute delays between the $\ell$-th Tx and Rx antennas and the $(n,m)$-th element of the RIS, respectively. Term $z_\ell(t)\in\mathcal{CN}(0,\sigma_z^2\delta_{\ell-k}\delta(t))$ is the additive noise, assumed as white and uncorrelated over different channels. The expression of $\rho$ follows from the radar equation \cite{skolnik}:
\begin{equation}\label{eq:rho}
    \rho = \sqrt{\frac{c^2}{(4\pi)^3 f^2_0 \|\mathbf{x}-\mathbf{x}_{\mathrm{T}}\|^2 \|\mathbf{x}_{\mathrm{R}}-\mathbf{x}\|^2} \Gamma_\mathrm{elem}} \;e^{j\delta}
\end{equation}
where $\Gamma_\mathrm{elem}$ is the radar cross section of the single RIS element of size $d^2$, and $\delta$ models residual phase uncertainties about the target (the RIS) arising, for instance, from Tx-Rx circuitry and Doppler effects from motion. We assume that Tx and Rx antennas, as well as RIS elements, are isotropic for simplicity, although more rigorous single-element models can be used \cite{Ellingson2021}. The overall path-loss depends only on the macroscopic Tx-RIS and RIS-Rx distances. 

Approximation $(a)$ in \eqref{eq:Rx_signal} assumes FF at both RIS and sensing terminal, i.e, uniform planar wavefronts, thus the delays can be linearized as
\begin{align}
    \tau^{i}_{\ell,nm} &\overset{(a)}{\simeq} \underbrace{\frac{\|\mathbf{x}-\mathbf{x}_{\mathrm{T}}\|}{c}}_{\tau^i_0} + \underbrace{\frac{\mathbf{x}_{\mathrm{T},\ell}^T \mathbf{u}(\boldsymbol{\zeta}_\mathrm{T})}{c}}_{\Delta \tau^i_\ell} + \underbrace{\frac{\mathbf{p}^T_{nm}\mathbf{u}(\boldsymbol{\xi}_i) }{c}}_{\Delta \tau^i_{nm}} \label{eq:delay_inc_approx}\\
    \tau^{o}_{\ell,nm} &\overset{(a)}{\simeq} \underbrace{\frac{\|\mathbf{x}_{\mathrm{R}}-\mathbf{x}\|}{c}}_{\tau^o_0} + \underbrace{\frac{\mathbf{x}_{\mathrm{R},\ell}^T \mathbf{u}(\boldsymbol{\zeta}_\mathrm{R})}{c}}_{\Delta \tau^o_\ell} + \underbrace{\frac{\mathbf{p}^T_{nm}\mathbf{u}(\boldsymbol{\xi}_o) }{c}}_{\Delta \tau^o_{nm}},\label{eq:delay_out_approx}
\end{align}
where $\tau^{i}_0$ and $\tau^{o}_0$ are the macroscopic delays between the Tx and Rx phase centers and the RIS phase center, respectively, while $\Delta \tau^{i}_\ell$ and $\Delta \tau^{o}_\ell$ are the excess delays at the sensing terminal, function of angles of departure and arrivals $\boldsymbol{\zeta}_\mathrm{T}$ and $\boldsymbol{\zeta}_\mathrm{R}$ respectively, whereas $\Delta \tau^{i}_{nm}$ and $\Delta \tau^{o}_{nm}$ are the excess delays at the RIS, function of incidence and reflection angles $\boldsymbol{\xi}_{i}=[\phi_i,\theta_i]^T$ and $\boldsymbol{\xi}_{o}=[\phi_o,\theta_o]^T$ (elevation and azimuth). The latter angles are computed as:
\begin{equation}
    \boldsymbol{\xi}_i =  J\left(\mathbf{Q}(\boldsymbol{\gamma}) (\mathbf{x}_\mathrm{T}-\mathbf{x})\right), \,\,\, \boldsymbol{\xi}_o =  J\left(\mathbf{Q}(\boldsymbol{\gamma}) (\mathbf{x}_\mathrm{R}-\mathbf{x})\right)
\end{equation}
where $J:\mathbb{R}^{3\times 1} \rightarrow \mathbb{R}_{2\pi}^{2\times 1}$ is the transformation between global Cartesian coordinates to RIS-local spherical coordinates, according to the convention in Fig. \ref{fig:system_model}. Unit vector $\mathbf{u}(\cdot)\in\mathbb{R}^{3\times 1}$ 
defines the direction in local or global coordinates.
For the planar RIS considered in this work, the excess delays are: 
\begin{align}
    \Delta \tau^{i}_{nm} & = \frac{d}{c}\left(n \sin\phi_{i} \cos\theta_{i} + m \sin\phi_{i} \sin\theta_{i}\right)\\
    \Delta \tau^{o}_{nm} & = \frac{d}{c}\left(n \sin\phi_{o} \cos\theta_{o} + m \sin\phi_{o} \sin\theta_{o}\right).
\end{align}
The further approximation $(b)$ in \eqref{eq:Rx_signal} assumes that 
\begin{equation}\label{eq:narrowband_condition}
    g\left(t\hspace{-0.1cm}- \hspace{-0.1cm}\tau^i_0\hspace{-0.1cm}-\hspace{-0.1cm}\tau^o_0\hspace{-0.1cm} - \hspace{-0.1cm}\Delta\tau^i_{\ell}\hspace{-0.1cm} - \hspace{-0.1cm}\Delta\tau^o_{\ell} \hspace{-0.1cm}- \hspace{-0.1cm}\Delta\tau^i_{nm}\hspace{-0.1cm}-\hspace{-0.1cm}\Delta\tau^o_{nm}\right) \hspace{-0.1cm}\overset{(b)}{\approx}\hspace{-0.1cm} g\left(t\hspace{-0.1cm}- \hspace{-0.1cm}\tau^i_0\hspace{-0.1cm}-\hspace{-0.1cm}\tau^o_0\right)
\end{equation}
i.e., the base-band Tx waveform is not affected by the residual (excess) delays at the RIS and at the sensing terminal. The narrowband operation described by \eqref{eq:narrowband_condition} is currently assumed in all the literature \cite{Zhang2023_targetRIS,10001209} but does not practically match the reality in most of sensing acquisitions, where the employed bandwidth $B$ is large enough to induce \textit{reflection beam squinting} from the RIS (i.e., a frequency-dependent RIS reflection coefficient). To gain insight on the latter phenomenon, we can analyze the Rx signal in the frequency domain, as reported in \eqref{eq:Rx_signal_frequency}. Here, $G(f)$ is the Fourier transform of $g(t)$,
\begin{equation}
    \beta(f|\mathbf{\Phi}) = \rho \,\sum_{n,m} e^{j\Phi_{nm}} e^{-j 2 \pi (f_0+f) (\Delta\tau^i_{nm} +\Delta\tau^o_{nm}) }
\end{equation}
is the overall frequency-dependent RIS reflection coefficient (including path-loss) depending on the phase configuration $\mathbf{\Phi}\in\mathbb{C}^{NM\times 1}$, and $Z_\ell(f)\in\mathcal{CN}(0,N_0\delta_{\ell-k}\delta(f))$ is the noise in the frequency domain, with power spectral density $N_0$. The base-band frequency $f\in[-B/2,B/2]$ models the wideband response of both Tx/Rx arrays and the RIS (e.g., see \cite{Ye2018widebandBF} for further details) through $\beta(f|\boldsymbol{\Phi})$, that \textit{filters} the impinging signal $G(f)$. Notice that the narrowband approximation in the frequency domain implies that the RIS has the same reflection coefficient for all the frequencies $f$ (corresponding to $f_0$). In the following, we derive and compare the CRB for the model \eqref{eq:Rx_signal_frequency}, highlighting the wideband effects. 

\section{CRB calculation}\label{sect:CRB}
The model \eqref{eq:Rx_signal_frequency} for the $\ell$-th measurement channel can be stacked into a $L \times 1$ vector:
\begin{equation}\label{eq:Rx_signal_frequency_vector}
    \mathbf{y}(f) = \mathbf{a}(f,\boldsymbol{\theta}|\overline{\mathbf{\Phi}})  + \mathbf{z}(f)
\end{equation}
where $\mathbf{a}(f,\boldsymbol{\theta}|\overline{\mathbf{\Phi}})$ is the non-linear model relating the parameters to be estimated $\boldsymbol{\theta}=[\mathbf{x}^T, \boldsymbol{\gamma}^T]^T \in \mathbb{R}^{6 \times 1}$ with the observation, conditioned to the specific phase configuration of the RIS $\overline{\mathbf{\Phi}}$, while $\mathbf{z}(f)\sim\mathcal{CN}(\mathbf{0}, N_0 \mathbf{I}_L \delta(f))$. Notice that we assumed the knowledge of the reflection coefficient (including path-loss) $\rho$, that allows obtaining an optimistic lower bound on pose estimation. If $\rho$ is unknown, its real and imaginary parts can be included in the parameters to be estimated. The RIS can be configured according to: 
\begin{equation}\label{eq:RIS_config}
    [\overline{\mathbf{\Phi}}]_{nm} \hspace{-0.1cm}=\hspace{-0.1cm} \begin{dcases}
         2 \pi f_0 \left[\frac{\|\mathbf{x}_{nm}-\mathbf{x}_\mathrm{T}\|}{c} + \frac{\|\mathbf{x}_\mathrm{R}-\mathbf{x}_{nm}\|}{c}\right] & \text{NF} \\
         2 \pi f_0 \left[\frac{\mathbf{p}^T_{nm}\mathbf{u}(\boldsymbol{\xi}_i) }{c} + \frac{\mathbf{p}^T_{nm}\mathbf{u}(\boldsymbol{\xi}_o) }{c}\right] & \text{FF}
    \end{dcases}
\end{equation}
where in the first case (NF) the RIS is able to perfectly focus the impinging radiation towards the phase center of the sensing terminal, while in the second case (FF) the RIS is configured to reflect the signal from $\boldsymbol{\xi}_i$ to $\boldsymbol{\xi}_o$. The perfect RIS configuration represents an upper performance bound modeling the situation in which the RIS follows some focusing alignment procedure such as the one in \cite{9508872}, not necessarily implying the perfect knowledge of $\mathbf{x}$ and $\boldsymbol{\gamma}$ (and thus of $\mathbf{x}_{nm}$), which is the objective of the estimation. A similar consideration can be made for FF configuration. In any case, the NF/FF phase configuration is optimal only at $f_0$. 

Based on model \eqref{eq:Rx_signal_frequency_vector}, the Fisher information matrix (FIM) $\mathbf{F}$ is block-partitioned as follows:
\begin{equation}\label{eq:FIM_CRB}
    \mathbf{F} = \begin{bmatrix}
        \mathbf{F}_{\mathbf{x}\mathbf{x}}& 
        \mathbf{F}_{\mathbf{x}\boldsymbol{\gamma}}\\
        \mathbf{F}_{\boldsymbol{\gamma}\mathbf{x}}& \mathbf{F}_{\boldsymbol{\gamma}\boldsymbol{\gamma}} 
    \end{bmatrix}
\end{equation}
with straightforward dimensions, whose entries are
\begin{align}
\mathbf{F}_{\boldsymbol{\mu}\boldsymbol{\nu}} \hspace{-0.1cm}= \hspace{-0.1cm} \frac{2}{N_0} \Re \left\{ \int\limits_{-B/2}^{B/2} \left(\frac{\partial \mathbf{a}(f,\boldsymbol{\theta}|\overline{\mathbf{\Phi}})}{\partial \boldsymbol{\mu}}\right)^H \frac{\partial \mathbf{a}(f,\boldsymbol{\theta}|\overline{\mathbf{\Phi}})}{\partial  \boldsymbol{\nu}} df \right\}
\end{align}
where $\boldsymbol{\mu}$ and $\boldsymbol{\nu}$ can be any set of the parameters to be estimated, i.e., $\mathbf{x}$, $\boldsymbol{\gamma}$. Single FIM entries can be evaluated as in \eqref{eq:partial_a_partial_x}-\eqref{eq:partial_a_partial_gamma}, where $\boldsymbol{\tau}^i_{nm} = [\tau^i_{1,nm},...,\tau^i_{L,nm}]^T$, $\boldsymbol{\tau}^o_{nm} = [\tau^o_{1,nm},...,\tau^o_{L,nm}]^T$ are the delay vectors, whose gradient w.r.t. $\mathbf{x}$ and $\boldsymbol{\gamma}$ is computed as (exemplary for $\boldsymbol{\tau}^i_{nm}$):
\begin{figure*}
\begin{align}
    \frac{\partial \mathbf{a}(f,\boldsymbol{\theta})}{\partial \mathbf{x}} & =  \rho\, G(f) \sum_{n,m} e^{j\overline{\Phi}_{nm}} [-j2\pi(f_0+f)] e^{-j 2 \pi (f_0+f) (\boldsymbol{\tau}^i_{nm} + \boldsymbol{\tau}^o_{nm}) } \frac{\partial}{\partial \mathbf{x}} \left(\boldsymbol{\tau}^i_{nm} + \boldsymbol{\tau}^o_{nm}\right)  + \nonumber\\
    & + G(f) \frac{\partial \rho}{\partial \mathbf{x}}  \sum_{n,m} e^{j\overline{\Phi}_{nm}} e^{-j 2 \pi (f_0+f) (\boldsymbol{\tau}^i_{nm} + \boldsymbol{\tau}^o_{nm}) } \in \mathbb{C}^{L \times 3}\label{eq:partial_a_partial_x}\\
    \frac{\partial \mathbf{a}(f,\boldsymbol{\theta})}{\partial \boldsymbol{\gamma}}  & =\rho\, G(f) \sum_{n,m} e^{j\overline{\Phi}_{nm}} [-j2\pi(f_0+f)] e^{-j 2 \pi (f_0+f) (\boldsymbol{\tau}^i_{nm} + \boldsymbol{\tau}^o_{nm}) } \frac{\partial}{\partial \boldsymbol{\gamma}} \left(\boldsymbol{\tau}^i_{nm} + \boldsymbol{\tau}^o_{nm}\right) \in \mathbb{C}^{L \times 3}\label{eq:partial_a_partial_gamma}
\end{align}
\hrulefill
\end{figure*}
\begin{align}
    \left[\frac{\partial \boldsymbol{\tau}^i_{nm}}{\partial \mathbf{x}}\right]_\ell & = \frac{1}{c} \frac{(\mathbf{x}_{nm} - \mathbf{x}_{\mathrm{T},\ell})^T}{\|\mathbf{x}_{nm} - \mathbf{x}_{\mathrm{T},\ell}\|} \\
    \left[\frac{\partial \boldsymbol{\tau}^i_{nm}}{\partial \boldsymbol{\gamma}}\right]_\ell & = \frac{1}{c} \frac{(\mathbf{x}_{nm} - \mathbf{x}_{\mathrm{T},\ell})^T}{\|\mathbf{x}_{nm} - \mathbf{x}_{\mathrm{T},\ell}\|} \frac{\partial \left(\mathbf{Q}(\boldsymbol{\gamma}) \mathbf{p}_{nm}\right)}{\partial \boldsymbol{\gamma}}.
\end{align}
The components of the FIM under the FF approximation (and possibly narrowband settings) follow from \eqref{eq:partial_a_partial_x}-\eqref{eq:partial_a_partial_gamma} under the approximation of the delays in \eqref{eq:delay_inc_approx}-\eqref{eq:delay_out_approx} and condition \eqref{eq:narrowband_condition}.

\section{Numerical Results}\label{sect:results}

\begin{figure}
    \centering
    \subfloat[][]{\includegraphics[width=0.95\columnwidth]{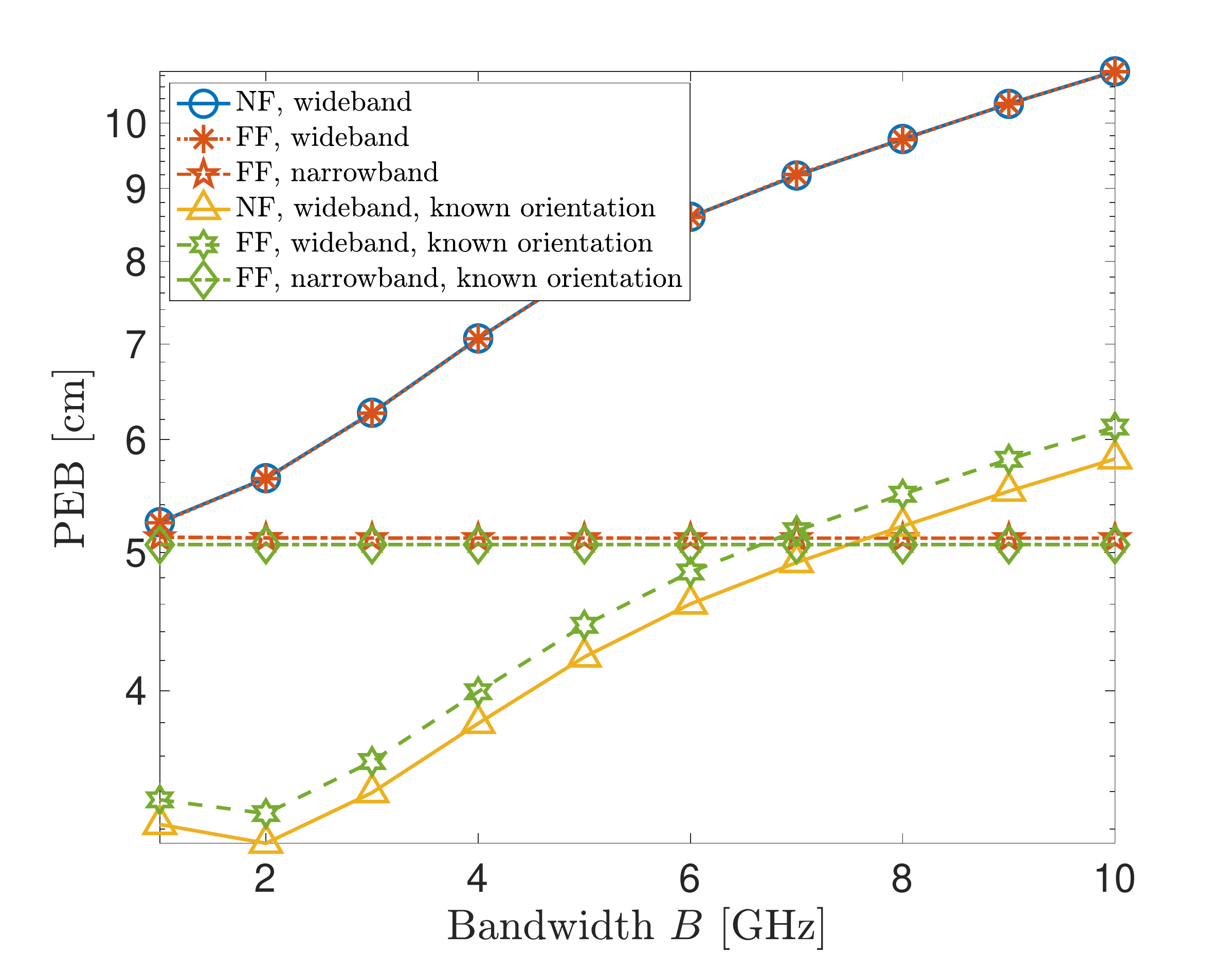}\label{subfig:PEB_vs_B_10cm2}}\vspace{-0.3cm}\\
    \subfloat[][]{\includegraphics[width=0.95\columnwidth]{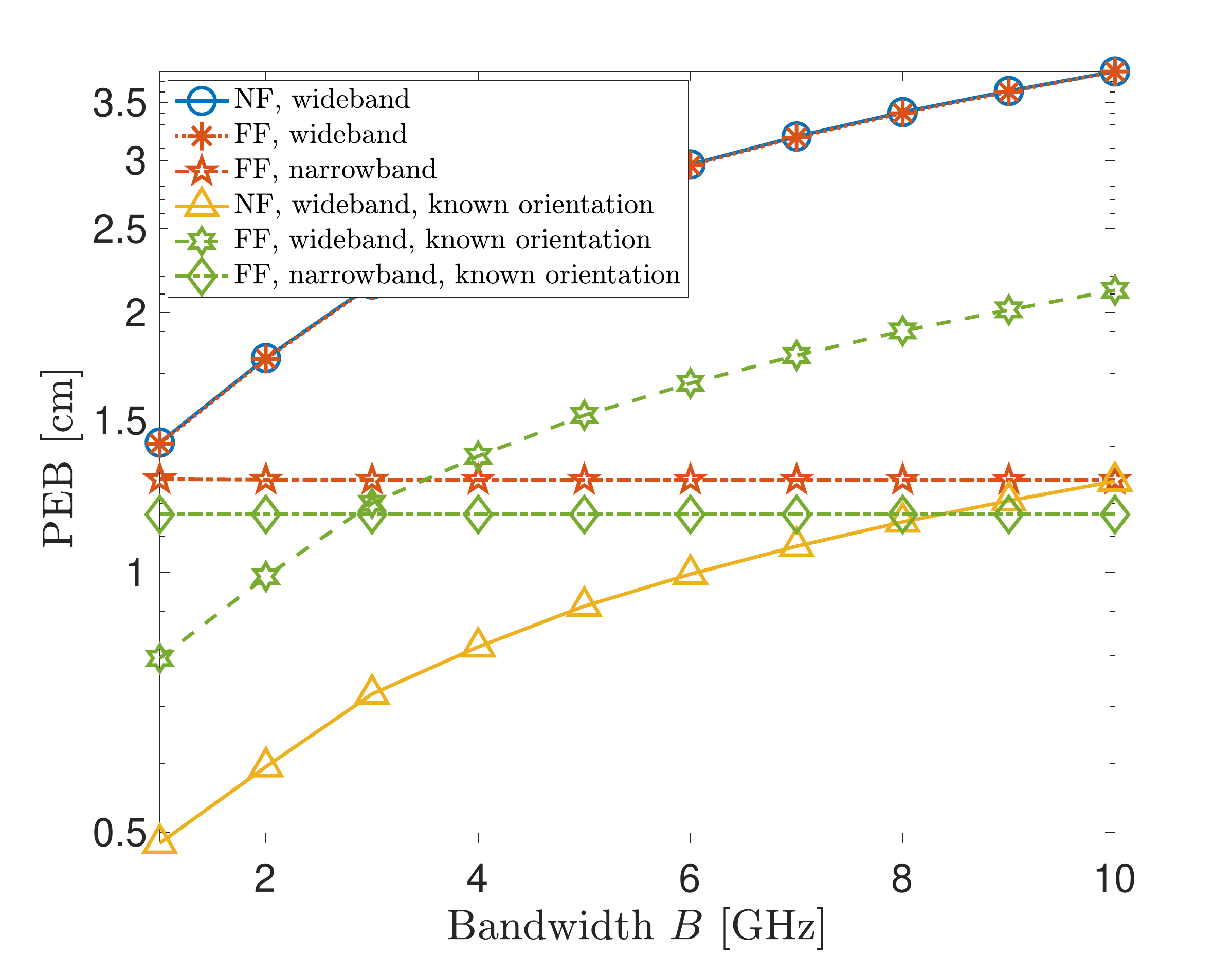}\label{subfig:PEB_vs_B_20cm2}}
    \caption{PEB vs. bandwidth $B$ for (a) $10 \times 10$ cm$^2$ RIS and (b) $20 \times 20$ cm$^2$ RIS. }
    \label{fig:PEB_vs_B}
\end{figure}
\begin{figure}
    \centering
    \includegraphics[width=0.95\columnwidth]{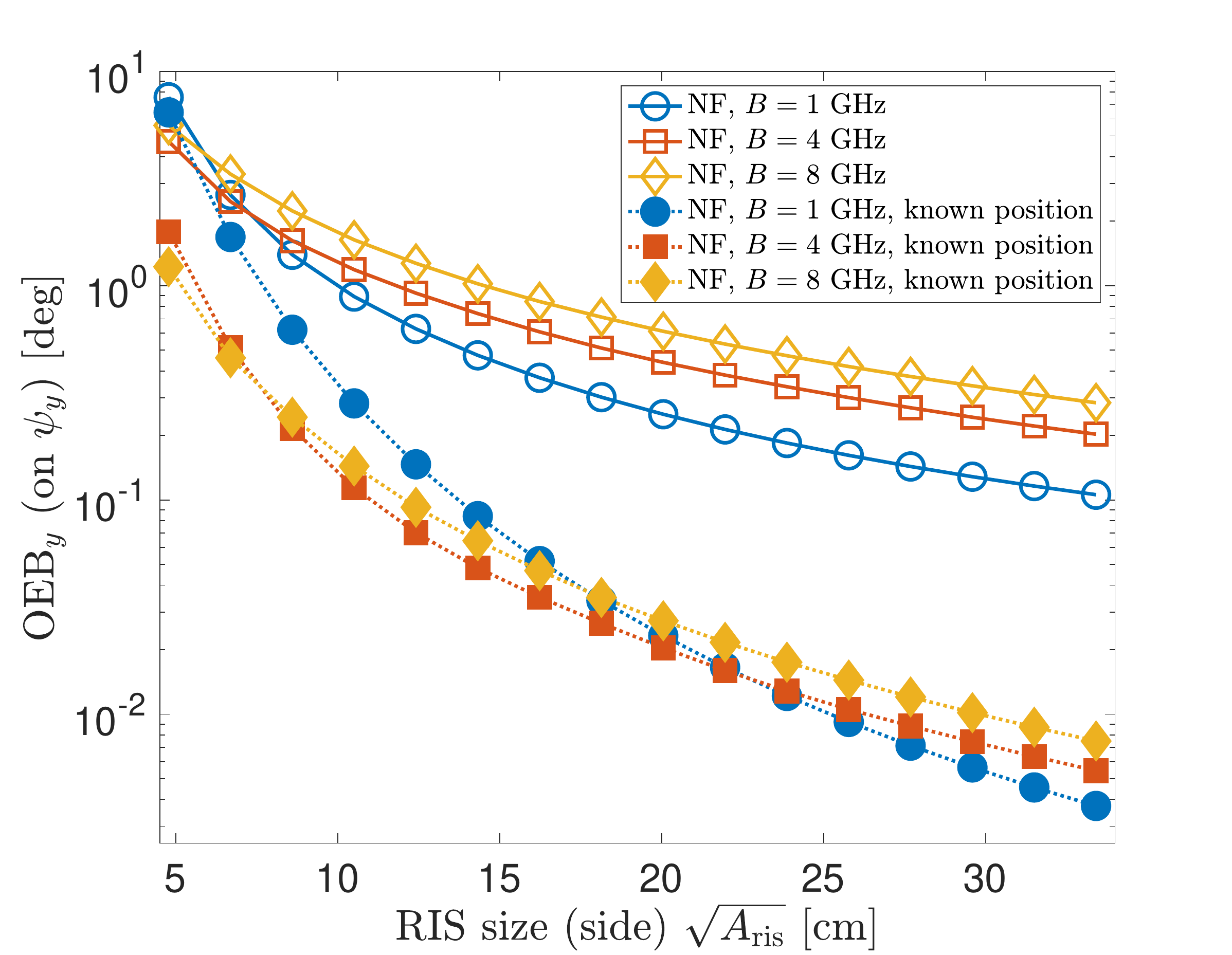}
    \caption{OEB of $\psi_y$ vs. RIS size (side size).}
    \label{fig:OEB_vs_L}
\end{figure}

This section shows numerical results quantifying the wideband effects for the pose estimation of a target-lodged RIS. We consider a monostatic sensing terminal operating at $f_0=78.5$ GHz, on a variable bandwidth $B\in[1,10]$ GHz, equipped with $1$ Tx antenna located in the origin of the global coordinate system and $20\times 20$ Rx antennas along $y$ and $z$ axes. The RIS size is variable, from $A_\mathrm{ris} = 5\times 5$ cm$^2$ to $A_\mathrm{ris} = 35\times 35$ cm$^2$. In all the evaluations, the RIS is in $\mathbf{x} = [5,0,-5.5]^T$ m, located on the $xy$ plane, that may represent a situation where the RIS is on a vehicle rooftop illuminated by a radar on a road side unit above. The Tx power is $23$ dBm at the single Tx antenna, and the Rx signal is corrupted by thermal noise, $N_0 = -173$ dBm/Hz. Notice that, as the power is kept fixed, increasing the bandwidth $B$ means decreasing the energy spectral density of the Tx signal $G(f)$.

The first set of results is summarized in Fig. \ref{fig:PEB_vs_B}. We show the position error bound (PEB), evaluated as 
\begin{align}
    \mathrm{PEB} = \sqrt{\frac{\mathrm{trace([\mathbf{F}^{-1}]_{1:3,1:3})}}{3}}\label{eq:PEB}
\end{align}
where the FIM can be either the NF or the FF one, under either wideband or narrowband operation. The latter, being the current state of the art \cite{Zhang2023_targetRIS,10001209}, assumes that the RIS behavior is independent on the base-band frequency $f$, as for approximation $(b)$ in \eqref{eq:Rx_signal} and \eqref{eq:Rx_signal_frequency}. We also report as a benchmark the lower bound in which the orientation $\boldsymbol{\gamma}$ is known (or its estimation is assumed to be decoupled from position), thus $\mathbf{F}=\mathbf{F}_\mathbf{xx}$. Let us first consider the case of a medium sized RIS, i.e., $10 \times 10$ cm$^2$ (Fig. \ref{subfig:PEB_vs_B_10cm2}). For increasing bandwidth $B$, the PEB increases as well, as long as the wideband modeling (both NF and FF) is concerned. PEB for narrowband FF is invariant, and it lower bounds the wideband one, underestimating the true PEB. The former effect is counter intuitive, as increasing $B$ is expected to decrease the PEB thanks to an increase in the sensing resolution. This not true, as the RIS in wideband conditions operates as a \textit{filter} on the Tx signal spectrum $G(f)$ that actually decreases the reflected energy w.r.t. the narrowband (frequency-flat) case. Notice that even when the RIS is assumed to be not frequency-selective (FF, narrowband curve), the PEB does not improve with $B$, as the effective bandwidth of $g(t)$ \eqref{eq:TX_signal} is \cite{BigS}:
\begin{equation}\label{eq:eff_band}
    B_\mathrm{eff}^2 = f_0^2 + \frac{B^2}{12} \approx f_0^2,
\end{equation}
largely dominated by the carrier component for the considered settings, and almost insensitive to $B$ (for $B=10$ GHz and $f_0=78.5$ GHz, the first term of \eqref{eq:eff_band} is 3 orders of magnitude higher than the second). Fig. \ref{subfig:PEB_vs_B_10cm2} also highlights the difference between the NF and FF modeling of the RIS behavior. When the orientation $\boldsymbol{\gamma}$ is known, e.g., a RIS is deployed on the rooftop of a vehicle, the FF shows higher PEB, as expected, as NF enables energy focusing at the Rx side.

Interestingly, for known $\boldsymbol{\gamma}$, the wideband PEB can be lower than the narrowband one. Thus, the information brought in the FIM  by the \textit{pose-dependent filtering} operated by the RIS overcomes the energy loss due to beam squinting and yields lower PEB. As $B$ increases, the energy loss becomes dominant and wideband PEB starts to be larger than the narrowband one (i.e,, the NF/FF wideband PEB curves cross the FF narrowband one). This latter threshold effect depends on the size of the RIS and on its distance from the sensing terminal. Although not reported for brevity, it can be shown that as the latter increases, the information loss due to spatial energy spreading dominates over the pose-dependent filtering effect, shifting the threshold towards lower values of $B$. This phenomenon is also observed by increasing the size of the RIS (see Fig. \ref{subfig:PEB_vs_B_20cm2}). In this latter case, the PEB gap between NF and FF---when $\boldsymbol{\gamma}$ is known---is higher, whereas again no difference between NF and FF is observed for the pose estimation problem. 

The second result, in Fig. \ref{fig:OEB_vs_L}, reports the orientation error bound (OEB) on the pitch angle $\psi_y$ for the NF model, versus the RIS size, varying the bandwidth $B = 1,4,8$ GHz. The OEB is defined as
\begin{equation}\label{eq:OEB}
    \mathrm{OEB}_y = \sqrt{[\mathbf{F}^{-1}]_{5,5}}
\end{equation}
where the lower bound corresponding to known position $\mathbf{x}$ is obtained by plugging $\mathbf{F}=\mathbf{F}_{\boldsymbol{\gamma}\boldsymbol{\gamma}}$ in \eqref{eq:OEB}. We select the pitch angle $\psi_y$ to be analyzed as it is the one providing the lower OEB among $\psi_x,\psi_y,\psi_z$, thus the higher model sensitivity. We notice that increasing the size of the RIS means increasing its frequency-selectivity for a given bandwidth $B$ (narrowing its reflection beamwidth and exacerbating the wideband effect), thus penalizing the Rx signal energy. The result is a OEB that worsens with $B$ except for very small RIS. For known $\mathbf{x}$, instead, this latter effect is only observed above a given RIS size, approximately $20\times 20$ cm$^2$ in the considered settings. Below, the lower OEB value is attained by $B=4$ GHz, as a trade-off between frequency-selectivity (dominating for large $B$) and a beneficial pose-dependent filtering  (dominating for small $B$). This is a further confirmation of what observed on the PEB in Fig. \ref{fig:PEB_vs_B}, and underlines the importance of the wideband modeling in the estimation of pose of a target-lodged RIS. As a final remark, the frequency-dependent RIS behavior opens for further research on the optimal spectral shaping of $G(f)$ to optimize (lower) the PEB/OEB (not covered here). 

\section{Conclusion}\label{sect:conclusion}
This letter analyzes the wideband effects on the pose (position+orientation) estimation of a target-lodged RIS, namely the frequency-dependent (filter) RIS behavior when the employed sensing signal bandwidth is large. We derive the CRB in NF/FF and wideband/narrowband modeling assumptions, discussing the trend of PEB and OEB varying the employed bandwidth and RIS size. The results highlight the importance of a wideband modeling of the RIS in the pose estimation problem, as both PEB and OEB worsen as the bandwidth increases. Under some circumstances and typical NF conditions, the wideband PEB is tighter than the narrowband one, as the RIS behaves like a pose-dependent filter.

\bibliographystyle{IEEEtran}
\bibliography{Bibliography}

\end{document}